# Efficient Solution of Language Equations Using Partitioned Representations

**Alan Mishchenko, Robert Brayton, Roland Jiang**
Department of EECS
University of California, Berkeley
Berkeley, CA 94720, USA
{alanmi, brayton, jiejiang}@eecs.berkeley.edu

**Tiziano Villa**
DIEGM
University of Udine
33100 Udine, Italy
villa@uniud.it

**Nina Yevtushenko**
Department of EECS
Tomsk State University
634050 Tomsk, Russia
ninayevtushenko@yahoo.com

**Abstract**

*A class of discrete event synthesis problems can be reduced to solving language equations $F \bullet X \subseteq S$, where $F$ is the fixed component and $S$ the specification. Sequential synthesis deals with FSMs when the automata for $F$ and $S$ are prefix closed, and are naturally represented by multi-level networks with latches. For this special case, we present an efficient computation, using partitioned representations, of the most general prefix-closed solution of the above class of language equations. The transition and the output relations of the FSMs for $F$ and $S$ in their partitioned form are represented by the sets of output and next state functions of the corresponding networks. Experimentally, we show that using partitioned representations is much faster than using monolithic representations, as well as applicable to larger problem instances.*

## 1 Introduction

Many important synthesis problems can be formulated in terms of solving language equations. The unknown $X$ represents an unknown component in a known network of components, which together must satisfy a given external specification. Examples are FSM synthesis, game solving, protocol conversion, discrete control, testing, etc. The solving for $X$ is typically very hard, involving operations that are exponential in complexity.

In many cases, hard computational problems can be reformulated using decomposition and partitioning, which lead to computational advantages. A good example is image computation [3], which is a core computation in formal verification. In its simplest form, the image of a set of states is computed using the formula:

$$\mathrm{Img}(ns) = \exists_{i,cs}[T(i,cs,ns) \wedge \xi(cs)],$$

where $T(i, cs, ns)$ is the transition relation, $\xi(cs)$ is the set of current states, $i$ is the set of input variables, and $cs$ ($ns$) is the set of current (next) state variables. The image, $\mathrm{Img}(ns)$, is the set of states reachable in one transition under all possible inputs from the current states, $\xi(cs)$, using the state transition structure given by $T(i, cs, ns)$.[1]

When the state transition structure is given by a sequential network, the transition relation can be represented, for example, in partitioned form using a set of next state functions, $\{T_k(i, cs)\}$, which update the state of each latch [2]. The partition, $\{T_k(i, cs, ns_k)\}$, represents how the next state of each latch, $ns_k$, depends on the values of inputs, $i$, and the current state, $cs$, of all latches:

$$T_k(i, cs, ns_k) = [ns_k \equiv T_k(i, cs)]$$

The complete transition relations can be derived as the product of the next state functions as follows:

$$T(i, cs, ns) = \Pi_k T_k(i, cs, ns_k) = \Pi_k [\, ns_k \equiv T_k(i, cs) \,]$$

We call $T(i, cs, ns)$ the *monolithic* representation of the transition relation and contrast it with the representation in terms of the partition, $\{T_k(i, cs, ns_k)\}$. Many problems can be solved using a partitioned representation without ever resorting to monolithic one, which may be impossible for larger problems. For example, the image computation can be performed using the partitioned representation by scheduling those $cs$ variables, which do not appear in some parts, to be quantified earlier [4][5]. In practice, this approach to image computation leads to significantly smaller intermediate BDDs, and has been responsible for dramatic increases, both in efficiency and in the sizes of problems that formal verification can solve.

The contribution of this paper is in proposing a way to use the partitioned representation for solving language equations of the type $F \bullet X \subseteq S$, for the case when the automata for $F$ and $S$ are prefix closed and represented by multi-level sequential networks.[3] We show how the partitioned representation can be used to perform the basic operations of language equation solving: completing, complementing, determinizing, hiding, and computing the product of finite automata.[4] An important aspect is that we show how most of these operations can be reformulated in terms of image computations, and hence can take advantage of the efficient developments in this area.

Section 2 gives the details of the problem statement for the specific case when the partitioned representation can be used. Section 3 outlines the computational procedures. Section 4 lists experimental results. Section 5 concludes. In the appendix, we prove that the determinization and completion operations commute, which makes the computations more efficient.

## 2 Problem Statement

Given a sequential network and its specification, it is theoretically possible to compute the Complete Sequential Flexibility (CSF) for a sub-part. The CSF represents all possible legitimate sequential FSM behaviors. They are legitimate in the sense that each can be implemented in a circuit and used to replace the selected sub-part, resulting in another network that satisfies the specification[5]. The initial research on capturing the maximum set of sequential behaviors was developed in [9].

---

[1] We do not distinguish between sets, relations, and their characteristic functions. Therefore, it is possible to think of $\xi(cs)$, $\mathrm{Img}(ns)$, and $T(i, cs, ns)$ as completely specified Boolean functions represented using e.g. BDDs.
[2] In this paper, we will illustrate the concepts using next state *functions*, but the computations also work for the case of non-deterministic relations.
[3] We assume the reader to be somewhat familiar with the concepts related to solving language equations (*e.g. see* [1]).
[4] Automata are used since all operations on FSMs are done by first converting these into automata by not distinguishing between inputs and outputs – a simple syntactic change. All states of the resulting automata are accepting states.

It can be shown [1] that the CSF, $X$, can be computed from the most general solution of the language equation, $F \bullet X \subseteq S$, where $F$ is the behavior of network without the selected part, $X$, and $S$ is the behavior of the original network as a whole or an external specification. The CSF is the largest prefix-closed, input-progressive automaton contained in $X$ (and thus an FSM).

The interaction of $F$ and $X$ within the specification $S$ is shown in Figure 1.[6] The external input and output variables are represented by symbols $i$ and $o$, respectively. The internal variables that are the inputs and outputs of $X$ are represented by symbols $u$ and $v$, respectively. Note that in Figure 1 and in the formulas below we do not distinguish between the sets of symbols used to construct the input/output languages of $F$, $X$, and $S$, and the sets of variables $i$, $u$, $v$, and $o$, used to encode these symbols.

The most general (the largest) solution of the language equation is the following [1]:

$$X = \overline{[F(i,v,u,o) \bullet (\overline{S(i,o)} \Uparrow^{u,v})]}_{\Downarrow u,v} \quad (1)$$

where symbol $\bullet$ denotes the composition of languages (the direct product of automata), the horizontal bars denote the complementation of a language (automaton), and the symbols $\Uparrow \Downarrow$ denote the expansion and restriction, respectively, of a language to a set of symbols. Expansion causes the insertion of the listed symbols into the language, while restriction leads to "hiding" the unlisted symbols on the inputs or arcs of the automaton.

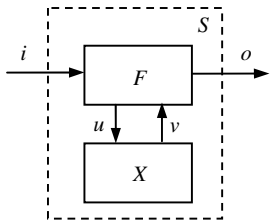

**Figure 1.** *A topology showing the interaction of components F and X in the language solving problem.*

We assume that the behaviors of the components $F$ and $S$ are represented by multi-level sequential networks, similar to the one in Figure 2. The external variables of such networks are the primary inputs ($i$), the primary outputs ($o$), and the latches having the current state ($cs$) and next state ($ns$) variables. We assume that the latch next-state functions, $\{T_k(i, cs)\}$, and the primary-output functions, $\{O_j(i, cs)\}$, can be computed and stored as BDDs in terms of the primary inputs and the current state variables.

The automata for $F$ and $S$ are derived, from the multi-level networks representing them, simply by taking the set of inputs of these automata as the union of the sets of inputs and outputs of the corresponding network. In the sequel, we use the terms "inputs" and "outputs" meaning the inputs and outputs of the network. All reachable states of a network are the accepting states of the corresponding automaton, since $F$ and $S$ are FSMs and hence are prefix-closed. A "don't-care" (DC) state can be added to make an automaton *complete*. This is done by making all automaton input combinations, for which the behavior of a state is not defined, transition to *DC*, which becomes the only non-accepting state. Prefix-closed dictates that *DC* has a universal self-loop.

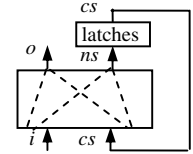

**Figure 2.** *The structure of a sequential network.*

*Example.* Figure 3 shows a sequential circuit and its corresponding automaton. The circuit has input $i$, output $o$, and two latches, with current state variables $cs = \{cs_1, cs_2\}$ and next state variables $ns = \{ns_1, ns_2\}$. The initial state of the latches is (00). The next state functions of the latches are $T_1(i, cs) = i \& cs_2$ and $T_2(i, cs) = \bar{i} + cs_1$. The two parts of the transition relation are

$$T_1(i, cs, ns_1) = [ns_1 \equiv T_1(i, cs)] = [ns_1 \equiv i \& cs_2] =$$
$$ns_1 \& i \& cs_2 + \overline{ns_1} \& (\bar{i} + \overline{cs_2})$$

$$T_2(i, cs, ns_2) = [ns_2 \equiv T_2(i, cs)] = [ns_2 \equiv (\bar{i} + cs_1)] =$$
$$ns_2 \& (\bar{i} + cs_1) + \overline{ns_2} \& i \& \overline{cs_1}).$$

The single output relation is $O(cs_1, cs_2, o) = [o \equiv (cs_1 \oplus cs_2)]$.

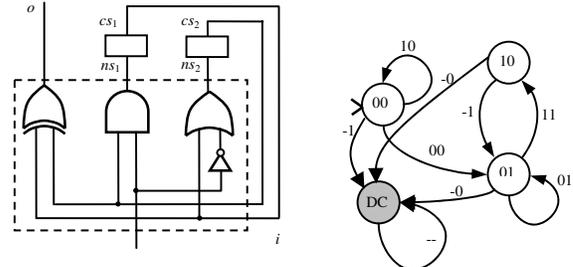

**Figure 3.** *A sequential network and its automaton.*

The states of the automaton in Figure 3 are labeled with the latch values ($cs_1,cs_2$). Transitions (arcs) are labeled with ($i,o$) values. Thus, the transition from state (00) under input 0 is to state (01). The output produced by the network in this case is 0. So the label of this transition is 00 (equal to 0/0 using conventional FSM labeling). This automaton is not complete. Thus, the transition from (00) under input (11) is not defined. The unshaded states are accepting. The additional (shaded) state (*DC*), added for completion, is not accepting. *DC* has a universal self-loop and all transitions that were originally undefined (e.g. from (00) under input (11)) are directed to *DC*.

**Monolithic Representation of Relations.** The monolithic representations of the transition and output relations of the automaton can be obtained from the next-state and output functions:

$$T(i, cs, ns) = \Pi_k [\, ns_k \equiv T_k(i, cs) \,],$$

$$O(i, cs, o) = \Pi_k [\, o_k \equiv O_k(i, cs) \,].$$

The monolithic representation of the complete transition-output relation of the automaton is:

---

[5] *However, it does not address implementations that can have combinational loops or that are not compositionally progressive* [10].
[6] *The results of this paper are not limited to the particular topology of Figure 1, but we confine the discussion to that of Figure 1 for ease of presentation.*

$$TO(i, o, cs, ns) = T(i, cs, ns) \& O(i, cs, o).$$

*TO* specifies what the next state is for each current state and each input/output combination. A relation is not well-defined if there exists some input/output/current-state combination for which the behavior of the automaton is not specified, i.e. the automaton is incomplete.

As an example of the use of the monolithic representation *TO*, consider the operation of completion of an incomplete automaton. The transition-output relation of the completed automaton is

$$TO'(I,o,cs,ns) = TO(i,o,cs,ns) + A(i,o,cs)DC(ns) + DC(cs)DC(ns),$$

where $A(i,o,cs) = \overline{\exists_{ns} TO(i,o,cs,ns)}$ is the sub-domain of input/output/current-state variables for which the original automaton is not defined, and *DC* is the code (in terms of latch variables) of the "don't-care" state added during completion. Note that we cannot use the code of an unreachable state to represent the *DC* because the unreachable states have next states. To encode the *DC* we need to add an additional state variable.

Thus the resulting relation is derived from the original one by simply directing all undefined transitions to *DC* and then adding a universal self-loop to *DC*.

**Partitioned Representation of Relations.** The disadvantage of monolithic representations is that they include all variables and hence their BDDs may be huge. Even if these can be computed, the operations, such as completion, product, determinization, become very inefficient if not impossible to complete. If the set of reachable states is much smaller than the set of all states, re-encoding the monolithic relations using fewer state bits may alleviate this problem. However, re-encoding can be very slow and our experience indicates that this tends to *increase* the BDD sizes of the relations.

In this paper, we show how partitioned representations can be used, both in making computations more efficient and in allowing larger problems to be solved. The monolithic transition and output relations are never constructed. With a partitioned representation, the functionalities of the automata are represented as sets of functions, $\{T_k(i, cs)\}$ and $\{O_k(i, cs)\}$. All Boolean operations are performed using these functions. It will be shown that image computation plays the key role in the manipulations involving partitioned representations, and hence a decade of research resulting in techniques to speed up image computations [4][5][8] can be used.[7]

## 3 Computation Algorithms

This section describes how to perform operations on the automata, required for language equation solving, when the partitioned representations are available. First, we present the main algorithm for the computation using either monolithic or partitioned representations. Later we show that, given the partitioned representations, all the steps are essentially embedded into a modified determinization procedure, so that there is no need to compute the completion and variable hiding operations as separate steps. However, in Section 3.2 for generality we show how these operations can be performed independently of determinization.

### 3.1 Main Algorithm

The main generic algorithm (shown below) first derives the most general solution from the equation $X = \overline{F \circ \overline{S}}$, followed by an additional prefix-closed and progressive operation on the solution *X*, to make it an FSM. Later, we show how to make the computations more efficient.

Algorithm: ***LanguageEquationSolving***
Input: prefix closed $S(i,o)$ and $F(i,v,u,o)$
Output: most general prefix closed solution *X*
**begin**
01   *X:=Complete(S)*
02   *X:=Determinize(X)*
03   *X:=Complement(X)*
04   *X:=Support(X,(i,v,u,o))*
05   *X:=Product(Complete(F),X)*
06   *X:=Support(X,(u,v))*
07   *X:=Determinize(X)*
08   *X:=Complete(X)*
09   *X:=Complement(X)*
10   *X:=PrefixClose(X)*
11   *X:=Progressive(X,u)*
12   return *X*
**end**

**Algorithm 1**: *Generic Algorithm for computing the most general prefix-closed progressive solution.*

*Support* (used for both expansion and restriction) changes its argument automaton to have the support of the given list of variables. *Complete* adds one additional non-accepting state with a universal self-loop, and causes all undefined inputs to transition to it. *PrefixClose* removes all non-accepting states. *Progressive* recursively removes those states that are not completely specified in terms of the input variables, *u*, of *X* (i.e. not input-progressive). The other operations are self-explanatory.

### 3.2 Elementary Operations using Partitioned Representations

In the application to the problem where the topology is as shown in Figure 1, we have partitions for *F* (which has both *u* and *o* as outputs):

$$\{T_j^F(i,v,cs_1,ns_j)\}, \{U_j(i,v,cs_1,u_j)\}, \{O_j^F(i,v,cs_1,o_j)\},$$

and those for *S*:

$$\{T_j^S(i,cs_2,ns_j)\}, \{O_j^S(i,cs_2,o_j)\}.$$

**Completion.** An automaton derived from an FSM is completed by interpreting the complement of its output relation as the condition for a transition from the current state to the non-accepting *DC* state. The complement of the output relation, $\overline{O(i,o,cs)}$, gives, for each current state, *cs*, all the input/output combinations, for which the automaton's behavior is not defined, since an FSM is defined for all its input combinations. When the partitioned representation is used, there is no need to construct the monolithic representation of $O(i,o,cs)$. For any current state, given by its characteristic function $\xi(cs)$, its undefined input/output combinations, $Q(i,o)$, can be found by an image computation followed by complementation:

---

$$Q(i,o) = \overline{\exists_{cs} O(i,o,cs) \& \xi(cs)}.$$

This image can be efficiently computed using the partitioned representation of the output relation:

$$Q(i,o) = \overline{\exists_{cs} \prod_j [o_j \equiv O(i,cs)] \& \xi(cs)}.$$

**Complementation (deterministic case).** In the general case of non-deterministic automata, determinization (subset construction) is required before complementation. In contrast, a deterministic automaton is easily complemented by interchanging the sets of its accepting and non-accepting states. Thus computing $\overline{S}$, which is deterministic, is easy.

When an automaton is derived from a deterministic multi-level network, its set of accepting states is equal to the set of reachable states of the network. The *DC* state introduced during completion is the only non-accepting state. In this case, complementation is performed by changing the interpretation of the *DC* state: it becomes the only accepting state, while all other states become non-accepting.

**Product computation.** The product of two automata is defined when they have the same support. In the partitioned view, having the same support simply means that each function is considered as a function of the full set of variables. When the argument automata are represented in their partitioned forms,

$$\{T_j^F(i,v,cs_1,ns_{1j})\} \text{ and } \{T_k^S(i,v,cs_2,ns_{2k})\},$$

the partitioned representation of the product automaton, is simply the union of the two partitions, i.e.

$$T(i,v,cs,ns) = \{T_{1j}(i,v,cs_1,ns_{1j}), T_{2k}(i,v,cs_2,ns_{2k})\}.$$

The fact that some variables, for example, $u$ and $o$, are missing, means that the relations are independent of these variables.

**Hiding variables.** Hiding variables $i$ and $o$ in the automaton representing the product of $F$ and $\overline{S}$ is an operation which changes the support automaton. In the monolithic form, it is performed by existentially quantifying the variables, not in the support, from the transition-output relation *TO* of the product automaton:

$$TO'(v,u,cs,ns) = \exists_{i,o}[\ TO(i,v,u,o,cs,ns)\ ].$$

This operation usually leads to a non-deterministic automaton, even if the original automaton is deterministic. Hiding variables cannot be performed on the partitioned representation because the operations of existential quantification and product do not commute. For example, it is not possible to quantify the input variables $i$ from the transition relation by quantifying them independently from each partition:

$$\exists_i T(i,cs,ns) = \exists_i \Pi_j [\ ns_j \equiv T_j(i,cs)] \neq \Pi_j \exists_i [\ ns_j \equiv T_j(i,cs)\ ].$$

Thus hiding can't be done by acting on the partitions independently. However, a key observation is that the hiding operation can be built into the next step, the determinization procedure, in such a way that there is no need to derive the monolithic transition relation and then apply hiding to it.

**Determinization (Subset Construction).** Determinization is performed by enumerating explicitly the subsets of states of a non-deterministic automaton, which are reachable from the initial subset state (which contains only the initial state). The basic step is the computation of the subset of states, reachable under various assignments of $(u,v)$ from a given subset of states, $\zeta(cs)$. Denote by $P_\zeta(u,v,ns)$ the sets of states (the subset states) reachable from $\zeta(cs)$ under various combinations of $(u,v)$. Using the monolithic representation, with variable $i$ hidden, the computation would be:

$$P_\zeta(u,v,ns) = \exists_{cs} \exists_i U(i,v,cs,u) T(i,v,cs,ns) \zeta(cs).$$

The computation of $P_\zeta$ using the partitioned form is:

$$P_\zeta(u,v,ns) = \exists_{i,cs} \Pi_j[u_j \equiv U_j(i,v,cs)] \Pi_k [ns_k \equiv T_k(i,v,cs)] \zeta(cs)$$

Thus, the computation of $P_\zeta(u,v,ns)$ can be seen as the computation of the image of $\zeta(cs)$, under the transition relation $U(i,v,cs,u) T(i,v,cs)$ represented in its partitioned form: $\{U_j(i,v,cs)\}$, $\{T_k(i,v,cs)\}$, where hiding (quantification) of variables $i$ is performed as part of the image computation.

Let $C_j(i,v,cs)$ denote the condition that is true when the $j$-th output of $F$ conforms to the $j$-th output of $S$:

$$C_j(i,v,cs) = \prod_k [O_{jk}^F(i,v,cs_1) \Rightarrow O_{jk}^S(i,cs_2)],$$

where $cs = (cs_1,cs_2)$ and the iteration is over all values $k$ of the output $O_j$, e.g. for binary signals, $k \in \{0,1\}$.

Let $C(i,v,cs) = \Pi_j C_j(i,v,cs)$, which is the overall condition when all outputs of $F$ conform to the outputs of $S$:

$$C(i,v,cs) = \prod_j \prod_k [O_{jk}^F(i,v,cs_1) \Rightarrow O_{jk}^S(i,cs_2)].$$

Next, we compute $Q_\zeta(u,v)$, the condition when the current subset of states $\zeta(cs)$ leads to the non-conformance of outputs of $F$ and $S$:

$$Q_\zeta(u,v) = \exists_{i,cs}[U(i,v,cs_1,u) \overline{C(i,v,cs)} \zeta(cs)].$$

Since these combinations of $(u,v)$ can lead to a non-conforming state, we exclude such transitions and map them all into a single non-accepting state *DCN* with a universal self loop. We remove these combinations from consideration by restricting $P_\zeta(u,v,ns)$ to those $(u,v)$ that are not contained in $Q_\zeta(u,v)$:

$$P_\zeta(u,v,ns) = P_\zeta(u,v,ns) \& \overline{Q_\zeta(u,v)}$$

Finally, the result is made complete by adding a state *DCA* (which is accepting in the *final* answer after complementation), with transitions from $\zeta(cs)$ into it for all $(u,v)$, which are not contained in $Q_\zeta(u,v)$.

Note that $\overline{C(i,v,cs)}$ can be represented as the sum of output non-conformance conditions for each output. Therefore, the computation of $Q_\zeta(u,v)$ can be done one output at a time, without computing the monolithic relation for $C(i,v,cs)$.

**Validity of the computation.** The maximum prefix-closed solution (required for an FSM implementation) is computed after determinization and completion, when the set of reachable subset states, $\{\zeta(cs)\}$, are known and, for each of these, the function $Q_\zeta(u,v)$ has been computed, which defines the transitions from this subset state into a newly added non-accepting state *DCA*.

To perform the complementation, which is the last step in computing the maximum solution, the accepting and non-accepting states are switched. To do this, we need to determine what the accepting states are after the previous steps of computation. It is helpful to follow the computation process starting from the initial automata, *F* and *S*, and consider the two types of their states: the accepting states (labeled *a*) and the non-accepting states.

- *S* could be completed by adding a new state $DC_1$ which would be its only non-accepting state, because *S* is represented by the multi-level network and so is prefix closed.

- In $\bar{S}$, $DC_1$ would be the only accepting state. (*If S is non-deterministic*, {$DC_1$} *would be the only accepting* (*subset*) *state*.)

- Since *F* was left incomplete and is prefix closed, all states of *F* are accepting.

- When forming the product $F \bullet \bar{S}$, a product state is accepting only if both states are accepting. Thus, in $F \bullet \bar{S}$, the only accepting states would be those with labels (*a*,$DC_1$) i.e. accepting in *F* and accepting in $\bar{S}$.

- In the subset construction, a subset state is accepting if at least one product state in the subset is accepting, i.e. of type (*a*,$DC_1$). If such occurs, then those transitions represented by $Q_\zeta(v,u)$ are redirected to *DCN,* since in the final answer, these states are not accepting[8]. The new completion state, *DCA*, added after the subset construction to complete it, is non-accepting for the determinized product $F \bullet \bar{S}$, but accepting in the final answer.

- Finally, in the complement of this product, which is the maximum solution, the set of accepting states are *DCA* plus all the subset states which are left in $P(u,v,cs)$ after $\overline{Q_\zeta}$ has deleted all those that lead to *DCN*.

Thus all non-accepting subset states are mapped into *DCN.* As mentioned, this can be done because we will make the answer *prefix-closed*, a requirement for an FSM. (Thus, the *X* computed is the most general *prefix-closed* solution.) This requirement increases efficiency, because during the subset construction, as soon as a state of type (*a*,$DC_1$) is encountered in a subset state, $\zeta$, it can be replaced with *DCN*. This efficiently reduces the number of subset states that emanate from such $S_k$. Further, those states that are trimmed immediately by this need not be explored for reachability from them. This leads to a substantial trimming during the subset construction. It can be traced back to the fact that *S* is prefix-closed, and we want *X* to be prefix-closed.[9]

Similarly, the existence of the single state, *DCA*, stems from the fact that *F* is prefix-closed. This allows for deferring the completion of *F* until the subset construction.

Finally, to make *X* into a FSM automaton, it is made input-progressive. This step is the same for both the monolithic and partitioned case and is not detailed here.

Note that in the partitioned flow, neither *S* nor *F* is made complete. In effect such completion is deferred to the all-inclusive determinization step. The validity of this is substantiated by the Appendix which proves that the determinization and completion operations commute and observes that completion also commutes with complementation and product.

## 4 Experimental results

In this section, an implementation of Algorithm 1 using the partitioned representations is compared with the implementation using the monolithic representations. In effect, the partitioned implementation performs only the determinization procedure, which subsumes all other steps, as described above. In the case of the monolithic representations, the completion of *S* is done first, then the intermediate product is derived, followed by hiding and determinization, performed in a traditional way.

The examples for this experiment were derived from FSM benchmarks by an operation called *latch splitting*. This is a syntactic transformation of a sequential circuit into two circuits, one containing a subset of the latches of the original circuit and the other containing the rest. One of these becomes the fixed component, *F*, in the language solving problem while the other represents a particular solution, $X_P$, for the "unknown" component. The sequential behavior of the original circuit is used as the specification. We compute the complete sequential flexibility (CSF – *the largest prefix-closed, input progressive sub-automaton of the most general solution.*) Since $X_P$, is a particular, it is contained in the CSF.

After computation of CSF, *X*, it was formally verified by checking

(1) $X_P \subseteq X$, and

(2) $F \bullet X \subseteq S \equiv F \bullet X_P$.

Table 1 lists the results for several examples. The columns show the example name (column "Name"), the number of inputs, outputs and latches (column "i/o/cs"), the number of latches in the fixed part and in the current implementation of the unknown (column "$F_{cs}/X_{cs}$"), the number of states in the CSF (column "States(*X*)"), the runtime in seconds of the partitioned algorithm (column "Part") and runtime in seconds using monolithic representations (column "Mono"), and finally the ratio of these runtimes (column "Ratio").

The algorithms were implemented in the MVSIS environment [6]. The measurements were made on a Windows XP computer with a 1.6 ghz CPU.

**Table 1.** *Comparison of partitioned and monolithic computations*.

| Name | i/o/cs | $F_{cs}/X_{cs}$ | States(*X*) | Part,s | Mono,s | Ratio |
|---|---|---|---|---|---|---|
| s510 | 19/7/6 | 3/3 | 54 | 0.3 | 0.2 | 0.7 |
| s208 | 10/1/8 | 4/4 | 497 | 0.4 | 0.8 | 2.0 |
| s298 | 3/6/14 | 7/7 | 553 | 0.9 | 2.7 | 3.0 |
| s349 | 9/11/15 | 5/10 | 2626 | 37.7 | 810.3 | 21.5 |
| s444 | 3/6/21 | 5/16 | 17730 | 25.9 | CNC | - |
| s526 | 3/6/21 | 5/16 | 141829 | 276.7 | CNC | - |

The results show that the partitioned method is more efficient (except for very small examples) with efficiency increasing as the problem size increases. At some point, the monolithic method can not complete (CNC) when the intermediate automata become relatively large. Of course, even the partition method runs out of steam at a certain point, but note that on one example it could handle over 140,000 states in the CSF.

---
[8] *Note that it is not necessary to compute subset states which emanate from such a state since once reached, we know that all input sequences with this prefix are not in the language of an FSM.*
[9] *While this trimming can be substantial, there is no avoiding that subset construction can be exponential. However, it has been a common experience among those who have implemented subset construction, that in practice, it can be surprisingly efficient with relatively few subset states reached, sometimes leading to a reduction in the number of states.*

## 5 Conclusions

We presented a strategy for solving language equations when the fixed component and the specification are represented using multi-level deterministic sequential networks. In such cases, a partitioned representation derived from the original network can be used to perform **all** the operations needed to compute the maximum solution. We showed the operations required can be re-formulated in terms of image computations, which can take advantage of the advances made in the efficiency of these computations over the last decade. The proposed methods allow for faster solution of larger problem instances. We emphasize that the techniques described in this paper rely heavily on the fact that both automata, $F$ and $S$, are represented initially by multi-level sequential networks and hence are prefix closed, and that the desired answer is also prefix closed.

In the Appendix, we argue that not completing $F$ in the algorithm leads to the same answer. This has two important consequences. First, it is not necessary to form the completion conditions before computing the product or its determinization. This saves a lot of computing. Second, completion of $F$ can lead to a non-deterministic product after hiding, while leaving $F$ incomplete leads to a deterministic product after hiding. Thus a subset construction is avoided when $F$ is not completed.

Finally, we note that finding the CSF is only one step in sequential synthesis. Finding an optimum sub-solution of the CSF remains the outstanding problem for future research.

## Acknowledgements

The first three authors gratefully acknowledge the support of the National Science Foundation under contract CCR-0312676, and the California Micro program with our industrial sponsors, Intel, Fujitsu, Magma, and Synplicity. The fourth author gratefully acknowledges the support of the MADESSII Project (Italian National Research Council). The fifth author was partly supported by the Russian Ministry of High Education (Grant UR.04.01.018). We gratefully acknowledge the support of a NATO travel grant (NATO Linkage Grant No. 971217).

## 6 Appendix: Validity of Non-Completion of $F$

**Theorem 1:** *To determinize an finite automaton A, the following two procedures are equivalent.*
1. Complete( Determinize( $A$ ), *non-accepting* )
2. Determinize( Complete( $A$, *non-accepting* ) )

**Proof**: We show that these two procedures yield the same result by comparing subset constructions without and with pre-completion. Let $DC$ denote the added non-accepting state. Recall that $DC$ has a universal self-loop. It is clear that if $\{s_1, ..., s_k\}$ is a subset state in the automaton constructed by Procedure 1, then there must exist subset state $\{s_1, ..., s_k\}$ and/or $\{s_1, ..., s_k, DC\}$ in the automaton constructed by Procedure 2. Also, if subset state $\{s_1, ..., s_k\}$ or $\{s_1, ..., s_k, DC\}$ exists in the resultant automaton of Procedure 2, then subset state $\{s_1, ..., s_k\}$ must exist in the resultant automaton of Procedure 1. Observe that

- $DC$ state is non-accepting. A subset state is accepting if and only if one of its states is accepting. Thus, state $\{s_1, ..., s_k\}$ is accepting if and only if state $\{s_1, ..., s_k, DC\}$ is accepting.
- Under any transition condition, the successor states of $\{s_1, ..., s_k\}$ are the same as those of $\{s_1, ..., s_k, DC\}$ except that the latter contains the $DC$ state.

Applying induction on the state space with the above two facts, it follows that subset states $\{s_1, ..., s_k\}$ and $\{s_1, ..., s_k, DC\}$ are equivalent. On the other hand, if a subset state $\{s_1, ..., s_k\}$ or $\{s_1, ..., s_k, DC\}$ can transition to the singleton subset state $\{DC\}$ under some transition condition, then all of the member states $s_1, ..., s_k$ must be incomplete under this transition condition. Therefore, it is immaterial whether the completion is done before or after the determinization.

QED

Other trivial propositions state that completion commutes with both the complementation and product operations. Corollary 1 follows from these observations and Theorem 1.

**Corollary 1:** *Let X be the solution obtained by Algorithm 1 when F and S are not completed and let Y be the solution when F and S are made complete first. Then the languages of X and Y are identical.*